# Title: Observation of High-Temperature Dissipationless Fractional Chern Insulator

**Authors:** Heonjoon Park[1], Weijie Li[1], Chaowei Hu[1], Christiano Beach[1], Miguel Gonçalves[2], Juan Felipe Mendez-Valderrama[2], Jonah Herzog-Arbeitman[2], Takashi Taniguchi[3], Kenji Watanabe[4], David Cobden[1], Liang Fu[5], B. Andrei Bernevig[2,6,7], Nicolas Regnault[8,2,9], Jiun-Haw Chu[1], Di Xiao[10,1], Xiaodong Xu[1,10*]

[1]Department of Physics, University of Washington, Seattle, Washington 98195, USA
[2]Department of Physics, Princeton University, Princeton, New Jersey 08544, USA
[3]Research Center for Materials Nanoarchitectonics, National Institute for Materials Science, 1-1 Namiki, Tsukuba 305-0044, Japan
[4]Research Center for Electronic and Optical Materials, National Institute for Materials Science, 1-1 Namiki, Tsukuba 305-0044, Japan
[5]Department of Physics, Massachusetts Institute of Technology, Cambridge, Massachusetts 02139, USA
[6]Donostia International Physics Center, P. Manuel de Lardizabal 4, 20018 Donostia-San Sebastian, Spain
[7]IKERBASQUE, Basque Foundation for Science, Bilbao, Spain
[8]Center for Computational Quantum Physics, Flatiron Institute, 162 5th Avenue, New York, NY 10010, USA
[9]Laboratoire de Physique de l'Ecole Normale Supérieure, ENS, Université PSL, CNRS, Sorbonne Université, Université Paris-Diderot, Sorbonne Paris Cité, 75005 Paris, France
[10]Department of Materials Science and Engineering, University of Washington, Seattle, Washington 98195, USA
*Corresponding author's email: xuxd@uw.edu

**Abstract:** The fractional quantum anomalous Hall effect has recently been experimentally observed in zero-field fractional Chern insulators (FCI). However, an outstanding challenge is the presence of a substantial longitudinal resistance $R_{xx}$ (a few kΩ), even though the anomalous Hall resistance $R_{xy}$ is quantized. This dissipative behavior is likely linked to imperfect sample quality. Here, we report transport measurements of a drastically improved twisted MoTe$_2$ bilayer device, which exhibits quantized $R_{xy}$ and vanishing $R_{xx}$ for the -2/3 state, marking a dissipationless FCI. Contrary to fractional quantum Hall states where the energy gap increases with magnetic field, we find that the thermal activation gap of the observed FCI states decreases rapidly as the magnetic field rises from zero, then plateaus above a few teslas. This observation is attributed to the interplay between spin and charge gaps. Due to the spontaneous ferromagnetism, the spin gap dominates at low field, while the charge gap becomes appreciable once magnetic field freezes spin fluctuations. For the -2/3 state, we estimate the spin and FCI gap of about 55 and 20 K, respectively. Our results gain insights into the energy scale of FCI and provide a pathway for quantum engineering of exotic correlated topological states.

**Main text:**

Fractional Chern insulators (FCIs)[1-6] are correlated topological phases of matter that exhibit fractionally quantized anomalous Hall effect (FQAH) in the absence of a magnetic field. They serve as lattice analogs of the celebrated fractional quantum Hall (FQH)[7,8] states observed in two-dimensional electron gas systems under strong magnetic fields. In both FQAH and FQH systems, the energy scale of fractional excitations is governed by electron-electron interactions, which are inversely proportional to the interaction length. For FQAH, this interaction length is set by the lattice scale λ, whereas in FQH systems, it is characterized by the magnetic length $l_B = 26$ nm/$\sqrt{B}$, where $B$ is the magnetic field in teslas. Since λ (typically a few angstroms) is smaller than experimentally achievable $l_B$ values, FQAH systems could in principle exhibit larger energy gaps compared to FQH systems. Recent progress has enabled the realization of FQAH in twisted MoTe$_2$ bilayers (tMoTe$_2$)[9-16] and rhombohedral moiré graphene[17-21]. In tMoTe$_2$, the -2/3 FCI state remains quantized at temperatures up to 1.5 K – approximately 40% of the material's Curie temperature[9]. Magnetization measurements[14] estimate a thermal activation gap of 2.3 meV (about 27 K) for this state, highlighting opportunities to further enhance the energy gap through improved sample quality and material engineering.

An intrinsic hallmark of the quantized Hall effect is dissipationless current flowing along sample edges, as evidenced by the quantized transverse resistance $R_{xy}$ and vanishing longitudinal resistance $R_{xx}$. While quantized $R_{xy}$ has been observed in FCI states of both tMoTe$_2$ and rhombohedral moiré graphene, $R_{xx}$ has remained finite, indicating residual dissipative transport. Drawing parallels from the development of FQH physics, this finite $R_{xx}$ likely reflects suboptimal sample quality. For instance, the initial observation of the 1/3 FQH state in GaAs quantum wells[7] – achieved in samples with modest electron mobilities of $10^5$ cm$^2$/(V·s) – showed quantized $R_{xy}$ (~$3h/e^2$) but finite $R_{xx}$ (~6 kΩ). Subsequent advances in sample quality led to mobilities surpassing $5\times10^7$ cm$^2$/(V·s), resulting in vanishing $R_{xx}$ and enabled the discovery of numerous FQH states[22-24]. A similar trajectory could be anticipated for FCIs, underscoring the importance of improving material quality and device fabrication.

Here, we report transport measurements on a refined tMoTe$_2$ device that uncovers several new facets of this FQAH system, including high temperature FCI states with vanishing $R_{xx}$. To achieve superior material quality, we employed horizontal flux growth (Methods). Conductive atomic force microscopy of the as-grown bulk MoTe$_2$ crystal reveals a defect density roughly two orders of magnitude lower than that of commercially available crystals used in the initial FQAH study[9] (Extended Data Fig. 1). Leveraging these crystals, we fabricated high-quality devices with lower moiré disorder by optimizing our atomic force microscopy cleaning technique, which enables us to expel sub-surface hydrocarbon contamination outside the Hall bar region (Extended Data Fig. 2e). Our device features a dual-gate architecture with patterned local contact gates[9] to minimize contact resistance (Fig. 1a, Extended Data Fig. 2). The homobilayer MoTe$_2$ is AA-stacked with a twist angle of approximately 3.7° (Fig. 1b).

**FQAH phase diagram**

We investigate the FQAH phase diagram via transport measurements conducted at a base temperature of 10 mK. The measurement scheme is illustrated in Fig. 1c (see Methods for details). Figure 1d shows the reflective magnetic circular dichroism (RMCD) signal as a function of the moiré filling factor $v$ (carriers per moiré unit cell), and the electric field ($D/\varepsilon_0$). The observed ferromagnetic phase diagram is consistent with prior reports[11,25]. Figures 1e and 1f present $R_{xx}$ and $-R_{xy}$ versus $v$ and $D/\varepsilon_0$. Here, $R_{xx}$ and $R_{xy}$ are symmetrized and antisymmetrized, respectively, at an applied magnetic field $\mu_0 H = \pm 200$ mT. This small field suppresses magnetic fluctuations during carrier density sweeps without altering the overall FQAH phase diagram (see Extended Data Fig. 3 for zero-field data). In the range of low electric fields, the data reveal either vanishing or reduced $R_{xx}$ together with large $R_{xy}$ at $v = -1, -2/3, -3/5$, and $-4/7$. These features correspond to integer ($v = -1$) and fractional Chern insulators states, respectively.

The magnetic field dependence further confirms the nature of these states. Figure 1g displays the fan diagram of $R_{xx}$ plotted against $\mu_0 H$ and carrier density $n$ (see Extended Data Fig. 4). The carrier densities at which $R_{xx}$ minima occur shift linearly with $\mu_0 H$. The overlaid dashed lines in Fig. 1g indicate the expected $n$-$\mu_0 H$ dependence from the Streda formula $C = \phi_o \frac{\partial n}{\partial B}$, where $C$ is the Chern number and $\phi_o$ is the magnetic flux quantum. The excellent agreement between these lines and the measured minima corroborates that the observed states are Chern insulators with $C = -1, -2/3, -3/5$, and $-4/7$.

These findings resemble those in Ref.[9] yet exhibit notable quantitative improvements attributable to the enhanced device quality. In Ref.[9], only the $-2/3$ FCI state was clearly resolved in the $R_{xx}$ phase diagram, spanning a density range of approximately $2.5 \times 10^{11}$ cm$^{-2}$. In the present device, additional $-3/5$ and $-4/7$ states appear, and the density range for the $-2/3$ plateau has been reduced by a factor two to about $1.25 \times 10^{11}$ cm$^{-2}$, indicating reduced disorder. Furthermore, we can now probe lower doping levels, uncovering highly insulating states at high electric fields ($D/\varepsilon_0 > 100$ mV/nm) at the commensurate fillings $v = -2/3, -1/2, -1/3$, and $-1/4$. Such behavior indicates the formation of charge-order on the triangular moiré superlattice in the layer-polarized regime[9-14,26,27].

**Dissipationless FCI**

As shown in Fig. 1e, $R_{xx}$ either vanishes or is strongly suppressed across the observed FCI states. To clarify this behavior, Fig. 2a plots $R_{xx}$ (red) and $-R_{xy}$ (blue) as a function of filling factor $v$ at $D/\varepsilon_0 = 0$. The data are symmetrized and antisymmetrized at $\mu_0 H = \pm 100$ mT. For $v$ between $-0.9$ and $-1.2$, $R_{xx}$ vanishes while $|R_{xy}|$ is quantized at $h/e^2$ ($h$ is Planck's constant and $e$ is the elementary charge), indicative of a quantum anomalous Hall (QAH) state. At $v = -2/3$, $R_{xx}$ also vanishes while $|R_{xy}|$ lies within 0.3% of $3h/2e^2$, demonstrating a dissipationless FCI state.

Figure 2b provides a closer look at $R_{xx}$ as the carrier density is swept through fractional fillings (see Extended Data Fig. 2 for the zero-field data). When $v$ slightly deviates from $-2/3$, $R_{xx}$ quickly jumps from zero to about 10-20 k$\Omega$, while $|R_{xy}|$ also decreases (Fig. 2a, inset) rather than increasing as often seen in the FQH states. This behavior near $v = -2/3$ appears consistently not only in tMoTe$_2$[9], but also in rhombohedral moiré graphene[17,18] and requires further theoretical investigation. At $v = -3/5$, a pronounced dip in $R_{xx}$ can be seen reaching ~3 k$\Omega$, with $|R_{xy}|$ close to 97% of $5h/3e^2$, indicating a $-3/5$ FCI state. Additional, weaker minima in $R_{xx}$ appear at $v = -4/7$ and $-5/9$, which we discuss further through temperature-dependent measurements. Near $v = -1/2$, $R_{xx} \approx 10$ k$\Omega$ and $|R_{xy}|$ varies linearly with $v$, approaching $2h/e^2$ at half-filling of the flat Chern band.

This behavior is in line with the earlier study[9] and matches expectations for an anomalous composite Fermi liquid state[28,29].

To confirm the anomalous Hall effect, we measured $R_{xx}$ and $R_{xy}$ while cycling the magnetic field at selected $v$. Figure 2c and 2d presents the results for $v = -1$ and $-2/3$, showing clear magnetic hysteresis and quantized $|R_{xy}|$ plateaus at $\frac{1}{|C|}\frac{h}{e^2}$, thus confirming a quantized anomalous Hall response. The coercive field exhibits a strong doping dependence, peaking at about 100 mT for the QAH state and decreasing to around 15 mT for the $-2/3$ FCI state, consistent with previous reports[9,11,25].

Fig. 2a also highlights several additional notable features. At $v = -3/2$, $R_{xx}$ peaks near 10 kΩ, while the anomalous Hall signal $R_{xy}$ vanishes. Over the low-field regime (Fig. 1g), the density at which $R_{xx}$ peaks shows little dependence on magnetic field, suggesting a topologically trivial charge density wave[15] at $v = -3/2$. Two broad resistive peaks appear near $v \approx -1.3$ and $-0.74$ with $|R_{xy}| \approx R_{xx}$. Although further study is needed to pinpoint their origin, these peaks likely mark phase-space boundaries on the higher-doping sides of the $v = -1$ QAH and $v = -2/3$ FCI states, respectively, as seen in the V-shaped features of the $v$-$D/\varepsilon_0$ phase diagram of Fig. 1e. A broad peak near $v \approx -0.85$ probably indicates the lower-doping boundary of the $v = -1$ state, consistent with our earlier work[9]. Between $v = -0.85$ and $-0.74$, both $R_{xx}$ and $|R_{xy}|$ decrease, with $R_{xx}$ remaining larger than $|R_{xy}|$. This behavior suggests an anomalous Hall metal phase[30], i.e., a spontaneous ferromagnetic (spin/valley-polarized) metal in a partially filled Chern band.

The emergence of additional FCI states also reveals subtle differences in their electric field-driven phase transitions. Focusing on the $-2/3$ FCI state (Fig.2e), it is noteworthy that $R_{xy}$ remains quantized over an extended range of electric field near zero. Once $D/\varepsilon_0$ surpasses a critical threshold, $R_{xx}$ quickly rises and remains high over a range of electric fields. This marks the transition from a layer-hybridized honeycomb lattice to the layer-polarized triangular lattice regime (see also Fig. 1e and Extended Data Fig. 5). Further increases in $D/\varepsilon_0$ eventually lead to a metallic state, suggesting the melting of the commensurate $-2/3$ charge-order state. In contrast, for the $-3/5$ FCI state (Fig. 2f), the $R_{xx}$ peak develops upon layer-polarization but drops quickly and enters a metallic phase (Fig. 1e), indicating a comparatively weaker charge-ordered phase than in the $-2/3$ state. These distinctions between $-2/3$ and $-3/5$ FCI states may help illuminate phase transitions between fractional Chern insulators and other competing many-body ground states.

**High Temperature FCIs**

We investigate the energy gaps of FCIs through temperature-dependent measurements. Figures 3a and 3b show $R_{xx}$ and $-R_{xy}$ as a function of temperature and filling factor, with the data symmetrized (for $R_{xx}$) and antisymmetrized (for $-R_{xy}$) at $\mu_0 H = \pm 150$ mT. Line cuts at selected temperatures, displayed in Figs. 3c, d, reveal that at $v = -1$ (the QAH state), $R_{xx}$ remains strongly suppressed (~5 kΩ) even at a temperature of 14 K, the upper limit of our dilution fridge system. The corresponding $|R_{xy}|$ is around 15 kΩ, implying a large Hall angle of $\tan(\theta_H) \approx 3$. The temperature dependence confirms a Jain sequence of FCI states, following $|\frac{p}{2p-1}|$ with $p = 2, 3, 4, 5$. These FCI states remain robust up to about 5 K, as indicated by a Hall angle of $\tan(\theta_H) \approx 1$.

We extract $R_{xx}$ and $R_{xy}$ for the identified -2/3 and -3/5 FCI states and plot them in Figs. 3e and 3f. For $v$ = -2/3, $R_{xx}$ vanishes while $|R_{xy}|$ stays quantized below about 2.5 K, persisting down to our base temperature of 10 mK — a change of two orders of magnitude in temperature. This behavior contrasts with rhombohedral pentalayer graphene, where $R_{xy}$ evolves from $\frac{3}{2}\frac{h}{e^2}$ at roughly 400 mk to $h/e^2$ at nearly 40 mK, which represents a transition from the FQAH to an extended quantum anomalous Hall phase[18]. In our twisted MoTe2 devices, none of the observed FCI states exhibit comparable temperature-driven transitions (see Extended Data Fig. 6).

To further investigate the energy gaps of the FCI states, we performed temperature-dependent measurements at selected magnetic fields. Figure 4a shows the measured $R_{xx}$ versus temperature and filling factor at external magnetic fields of $\mu_0 H$ = 0, 0.15, 0.5, and 2 T. Additional data at other magnetic fields are plotted in Extended Data Fig. 7. The dashed line in the zero-field panel represents the Curie temperature $T_C$ determined from RMCD measurements (Extended Data Figs. 8, 9). Although applying a magnetic field does not create new FCI states, it enhances existing $R_{xx}$ dips, allowing them to persist at higher temperatures. This stabilization is expected when the gap of the FCI states exceeds $T_C$.[31-33] As the magnetic field aligns spins near $T_C$, the FCI states become more robust at elevated temperatures.

We estimate energy gaps by fitting $R_{xx}(T)$ to an Arrhenius form $R_o e^{-\frac{\Delta}{2k_B T}}$, where $R_o$ is a fitting constant, $\Delta$ is the energy gap, and $k_B$ is Boltzmann constant. Here, we focus on the -2/3 state, as its vanishing $R_{xx}$ allows more reliable fits compared to other FCI states. Figures 4b and 4c show the extracted $R_{xx}$ versus $T$ and its Arrhenius plot at different magnetic fields, respectively. The estimated gap $\Delta$ depends strongly on the magnetic field. At zero magnetic field, $\Delta = (52 \pm 4)$ K; it decreases to ~30 K by ~1T, and saturates around ~20 K above ~4 T (Fig. 4d). The suppression of $\Delta$ with increasing magnetic field is distinct from the conventional FQHE, where $\Delta$ typically increases with magnetic field ($\propto \sqrt{B}$),[22,23,34] as the gap is inversely proportional to the magnetic length $l_B$. In contrast, for FQAH systems, the interaction length is set by the lattice scale and does not depend on the external magnetic field. As a result, intrinsic FQAH gaps should exhibit minimal field dependence – opposite to our observation.

A plausible explanation is that when $\Delta$ surpasses $T_C$, temperature dependent resistance measurements become sensitive to spin fluctuations and/or magnetic domain effects, which causes $R_{xx}(T)$ to increase rapidly near $T_C$. These fluctuations dominate the linear region of $\ln(R_{xx})$ versus $1/T$ in Arrhenius plots, leading to an overestimation of FCI gaps. When an external magnetic field is applied, it suppresses these spin fluctuations and domain effects, reducing the slope in the Arrhenius fits and thus lowering the extracted gap. At sufficiently high fields, where spins become fully aligned due to the system's large $g$-factor[35,36], fluctuation-driven contributions are eliminated, and the thermal activation gap saturates at approximately 20 K. We identify this as the intrinsic charge gap of the $v$ = -2/3 FCI state, consistent with local magnetometry estimates (~27 K) from nanoSQUID-on-tip sensors[14].

To quantitatively assess this scenario, we fit the data using Matthiesen's rule (See Methods and Extended Data 10). We model transport as being governed by two types of excitations, the FCI gap ($\Delta_{FCI}$) and the spin transport gap ($\Delta_S$). We find that $\Delta_{FCI}$ is remarkably stable, varying only within $\Delta_{FCI} \in [18.8 - 20.9]K$, consistent with the saturated gap value. In contrast, the spin gap exhibits greater fluctuations, ranging from $\Delta_S^0 \in [50.3 - 66.2]K$. This larger variation is expected, as extracting the spin gap relies on an accurate estimate of the $g$-factor (Extended Data 10). Nevertheless, we consistently find $\Delta_S^0 \approx 3\Delta_{FCI}$.

**Discussion**

Finally, we turn to the filling factor dependence of FCI gaps. Within the framework of composite fermion theory[37], the energy gaps of FQH states are approximately proportional to $1/(2p+1)$. However, a precise theoretical understanding of these gaps has proven highly nontrivial[38-40]. In particular, experimentally measured gaps tend to be smaller than theoretical estimates, possibly due to finite quantum well width and residual disorder[41-44]. Yet in ultra-high-quality quantum wells, recent data do confirm the predicted Jain-sequence gap scaling[45]. A natural question is whether similar behavior emerges for FCI states.

A central challenge in our system is that the -4/7 and -5/9 states exhibit substantial residual $R_{xx}$ at low temperatures (See Extended Data Fig. 6), complicating accurate gap extraction. Focusing on the -3/5 state, where $R_{xx} \approx 2.5$ k$\Omega$ is relatively small, yields an estimated gap $\Delta_{-3/5} \approx 6$ K at high fields – compared to $\Delta_{-2/3} \approx 20$ K (Fig. 4d). Their ratio (~3) modestly exceeds the naïve expectation (~5/3) based on $1/(2p+1)$. Going forward, improved devices and/or more direct gap probes (such as chemical potential measurements) will be crucial for clarifying whether a universal FCI gap-scaling relation emerges. As the history of FQH research demonstrates, material and device quality are critical for the realization of new states of quantum matter. Our results demonstrate the possibility of improving twisted MoTe$_2$, reinforcing its potential as a promising platform for exploring strongly correlated topological phenomena.

**Methods**

**MoTe$_2$ Crystal Growth and characterization.** The MoTe$_2$ crystals are grown with horizontal flux transport. Around 40 g of Mo and Te pieces with a molar ratio of 1:20 are loaded in a quartz ampule with a pre-made neck. The two sides of the neck are kept at 600 °C and 550 °C respectively during the growth. The process is illustrated in Extended Data 1a. Mo pieces are initially placed on the hot end, and in a month mm-sized crystals can be obtained on the cold end. After being separated from Te flux, the crystals are annealed overnight to remove excess Te in another quartz tube on the hot end at 380 °C while the cold end is kept near room temperature. Then MoTe$_2$ crystals are exfoliated onto a Si chip pre-coated with 2 nm Cr and 5 nm Au to be characterized by conductive AFM. The conductive AFM is performed on an Asylum Research Cypher ES instrument using a gold-coated tip manufactured by BudgetSensors with a spring constant of 0.2 N/m in an ambient environment. To maintain the tip quality during the scan, the scanning rate is kept under 2 $\mu$m/s and the bias voltage is kept low so that the maximal current is under 10 nA.

**Device fabrication.** The contact gate structure used in this study is fabricated using the multi-step dry-transfer technique detailed in a previous work[9]. Hexagonal boron nitride (hBN) and graphite (Gr) were exfoliated and the thickness were confirmed using optical microscopy and atomic force microscopy (AFM). A bottom gate structure was created using poly (bisphenol A carbonate)/polydimethylsiloxane stamp mounted on a glade slide by sequentially picking up hBN (6.7 nm) and Gr and melting down on a 90 nm SiO$_2$/Si substrate. Standard e-beam lithography techniques were used to define a Hall bar pattern on the bottom gate. Ti/Pt (2 nm/5 nm) contacts were evaporated and cleaned using contact-mode AFM. An MoTe$_2$ flake was exfoliated on a 285 nm SiO$_2$/Si substrate in a glovebox with O$_2$/H$_2$O levels less than 0.5 ppm and cut in half using an AFM tip. An hBN (3.2 nm) contact gate dielectric, one-half of the MoTe$_2$ flake was first picked up and the latter half was picked up after rotating the stage by 3.7 degrees to form the tMoTe$_2$ structure. After melting down on the bottom gate, the structure was AFM cleaned again to remove bubbles and check if any cracks or wrinkles have formed. A Ti/Pt (2/5 nm) contact gate was evaporated on top of the Hall bar pattern to decrease contact resistance and AFM cleaned again. Finally, the top gate structure consisting of hBN/Gr/hBN (4.3 nm) was melted down to finish the device. Cr/Au (7nm/120 nm) contacts were evaporated to connect the Pt contacts to outer pads for wire-bonding.

**Electrical measurements.** Transport measurements were conducted in a Bluefors dilution refrigerator with a 9 T magnet. Standard low-frequency lock-in techniques using SR830 and SR860 lock-ins were used to measure resistances $R_{xx}$ and $R_{xy}$. An excitation current of 0.2-0.5 nA was used with a frequency of 4.777 – 7.777 Hz.

**Reflective magnetic circular dichroism (RMCD) measurements.** Experiments were conducted in a closed-loop cryostat system (attoDRY 2100) with precise sample positioning via an xyz piezo stage, a 9T superconducting magnet oriented out-of-plane, and a base temperature of 1.6 K. The excitation source was a filtered broadband supercontinuum laser (NKT SuperK) passed through a monochromator twice to isolate a narrow spectral range aligned with the trion resonance. Reflectance differences arising from right- and left-circularly polarized light (RCP and LCP) were monitored to quantify the magneto-optical response.

Light intensity was modulated at 850 Hz using a chopper, and polarization alternated at 50 kHz through a photoelastic modulator, enabling precise differentiation between polarization states. Reflected light was captured by an InGaAs avalanche photodiode and analyzed using dual lock-in amplifiers (SR830) to simultaneously measure reflectance variation and total reflectance. The normalized RMCD signal was derived from the ratio of phase-modulated and intensity-modulated signals, incorporating a correction factor based on the first-order Bessel function.

**Estimation of filling factor based on doping density.** The carrier density ($n$) and electric field ($D/\varepsilon_0$) in the sample were calculated from the applied top and bottom gate voltages $V_{tg}$ and $V_{bg}$ using the relations: $n = (V_{tg}C_{tg} + V_{bg}C_{bg})/e - n_{offset}$ and $D/\varepsilon_0 = (V_{tg}C_{tg} - V_{bg}C_{bg})/2\varepsilon_0 - D_{offset}$. Here, $C_{tg}$ and $C_{bg}$ are the capacitances of the top and bottom gates determined by measuring the hBN thickness through atomic force microscopy, $e$ is the electron charge, and $\varepsilon_0$ is the vacuum permittivity. The offset values, $n_{offset}$ and $D_{offset}$, were extracted by aligning experimental data with characteristic features observed in the dual-gate resistance maps, such as the sharp resistive peaks observed at integer or fractional fillings.

**Two-gap fit for thermal activation of the longitudinal resistance.** As shown in Fig. 4c of the main text, the thermal activation at small and large magnetic field $B$ is set by different charge gaps that differ by more than a factor of 2, with the larger gap being observed at $B = 0$. This is in sharp contrast to what is observed for the FQHE, where the transport gap increases with $B$. A plausible interpretation is that thermal activation behavior at small $B$ is dominated by spin fluctuations, which are suppressed as $B$ increases as mentioned in the main text. In particular, at large enough $B$, these fluctuations would be completely suppressed, implying that the Arrhenius slope is determined by the (fully polarized) FCI quasi-particle charge gap. Thus, thermo-transport experiments can reveal the intrinsic FCI gap once the spin fluctuations are frozen by $B$ field.

To make this scenario quantitative, we propose a description of the activation behavior of the longitudinal resistance, $R_{xx}$, by assuming that transport is dominated by two types of charge excitations that carry opposite valley quantum numbers. These carriers are, respectively: i) a fractionalized quasiparticle on top of the ground state of the FCI, which constitutes the lowest energy charged excitation with a gap $\Delta_{FCI}$, and ii) an electronic quasiparticle carrying spin opposite that of fractionalized quasiparticles with a gap $\Delta_S$. Since the ground-state state is polarized it is natural to assume that $\Delta_{FCI} < \Delta_S$. We denote the two different types of carriers by $\gamma = FCI, S$ respectively and their gaps by $\Delta_\gamma$.

At temperatures below $T_C$, and once the FCI develops, the ground-state breaks time-reversal symmetry but preserves valley/spin $U(1)$ symmetry. If spin-flip scattering processes are negligible, carriers with opposite spin quantum numbers will constitute independent conduction channels that add in parallel. The simplest assumption is that each of the two carrier species has a conductivity given by the Drude formula $\sigma_{xx}^\gamma = n^\gamma \mu^\gamma e^\gamma$ with $n^\gamma$ being the density of the $\gamma$ carriers, $\mu^\gamma$ their individual mobilities, and $e^\gamma$ their charges. We then assume that the temperature dependence of the conductivities in the range of the experiment is dominated by the thermal activation of the density of these charge excitations with $n^\gamma \propto e^{-\frac{\Delta_\gamma}{2k_BT}}$, while the mobilities $\mu^\gamma$, which encode their respective scattering rates, remain constant. In this simple scenario, the conductivity of each carrier can be described by $\sigma_{xx}^\gamma = \sigma_{xx0}^\gamma e^{-\Delta_\gamma/2k_BT}$, with $\sigma_{xx0}^\gamma$ a temperature-independent constant. Under these assumptions, the total conductivity of the system is determined by

$$\sigma_{xx} = \sigma_{xx0}^S(e^{-\frac{\Delta_S(B)}{2k_BT}} + \alpha e^{-\frac{\Delta_{FCI}}{2k_BT}}) \quad (1)$$

where $\alpha = \sigma_{xx0}^{FCI}/\sigma_{xx0}^S$. In what follows, we assume that the FCI gap, $\Delta_{FCI}$, is independent of magnetic field, as suggested by the large $B$ saturation behavior observed in experiments for the transport gap (see Fig.4c,d). The spin gap, on the other hand, depends on $B$ through the Zeeman effect as $\Delta_S = \Delta_S^0 + 2g^*\mu_B B$, where $g^*$ is the effective g-factor renormalized by interactions, $\mu_B$ is the Bohr magneton, and $\Delta_S^0$ is the zero magnetic field spin gap. The longitudinal resistivity is then given by

$$\rho_{xx} = \frac{\sigma_{xx}}{\sigma_{xy}^2+\sigma_{xx}^2} \approx \frac{\sigma_{xx}}{\sigma_{xy}^2} = \frac{\sigma_S}{\sigma_{xy}^2}(e^{-\frac{\Delta_S(B)}{2k_BT}} + \alpha e^{-\frac{\Delta_{FCI}}{2k_BT}}) \quad (2)$$

where we assumed $\sigma_{xy}^2 \gg \sigma_{xx}^2$ which is justified in the low-$T$ thermal activation regime. The longitudinal resistance is then fit to the expression

$$R_{xx} = \tilde{R}_0(e^{-\frac{\Delta_S(B)}{2k_BT}} + \alpha e^{-\frac{\Delta_{FCI}}{2k_BT}}) \quad (3)$$

where $\tilde{R}_0$ is a fitting constant. We truncate the data to carry out the fitting procedure and focus on the regime in which clear thermal activation is observed below the high-temperature resistance saturation and for $0.15 \lesssim T^{-1} \lesssim 0.35 K^{-1}$. The resulting fit displays good agreement with the data in a wide range of temperatures and magnetic fields by introducing only five free parameters ($g^*, \Delta_{FCI}, \Delta_S^0, \tilde{R}_0$, and $\alpha$) as shown in Extended Data Fig.10. At lower temperatures, $T^{-1} \gtrsim 0.35 K^{-1}$, Eqn.3 no longer describes the observed temperature dependence of $R_{xx}$. These deviations from thermally activated transport may be explained by other mechanisms, such as variable-range hopping, becoming dominant[46,47]. However, a detailed understanding of this low-temperature regime is limited by the signal-to-noise ratio of $R_{xx}$ at $T^{-1} \gtrsim 0.35 K^{-1}$.

In order to inspect the robustness of the best-fit parameters, we tested several fitting protocols by varying the truncation of the data and implementing different weighting functions (see caption of Extended Data Fig.10). To avoid the high-temperature saturation regime and ensure we are in the regime of validity of the fit where $(R_{xx}/R_{xy})^2 \ll 1$, we limit the range of resistivities with a maximum value $R_{xx}^{max}$. In Extended Data Figs.10b-d, we show the fit parameters as a function of $R_{xx}^{max}$. For all fit parameters we see variations consistent within error-bars and a small systematic drift set by the weighting function (see caption of Extended Data Fig.10). The consistency of the extracted parameters confirms that the data is not being over-fitted. We find the FCI gap $\Delta_{FCI}$ is the most robust parameter, varying only between $\Delta_{FCI} \in [18.8 - 20.9]K$. The spin gap fluctuates more significantly between $\Delta_S^0 \in [50.3 - 66.2]K$. This larger variation is expected, as the extraction of the spin gap depends on an accurate estimate of the $g$-factor. However, we consistently find $\Delta_S^0 \approx 3\Delta_{FCI}$. Furthermore, the results for both gaps are consistent with the $B$−resolved Arrhenius fits in the main text.

An interesting observation from Eqn.3 is that, to explain the data, we require $g^* \in [17.4 - 26.4]$, which is significantly larger than previous theoretical estimates[35,36] and calls for further study. Finally, we note that the results for the parameter $\alpha$ imply that the spin channel is significantly more dissipative by 2 to 3 orders of magnitude.

**References:**
1. Neupert, T., Santos, L., Chamon, C. & Mudry, C. Fractional quantum Hall states at zero magnetic field. *Physical review letters* **106**, 236804 (2011).
2. Sheng, D., Gu, Z.-C., Sun, K. & Sheng, L. Fractional quantum Hall effect in the absence of Landau levels. *Nature communications* **2**, 389 (2011).
3. Regnault, N. & Bernevig, B. A. Fractional chern insulator. *Physical Review X* **1**, 021014 (2011).
4. Tang, E., Mei, J.-W. & Wen, X.-G. High-temperature fractional quantum Hall states. *Physical review letters* **106**, 236802 (2011).
5. Sun, K., Gu, Z., Katsura, H. & Das Sarma, S. Nearly flatbands with nontrivial topology. *Physical review letters* **106**, 236803 (2011).
6. Xiao, D., Zhu, W., Ran, Y., Nagaosa, N. & Okamoto, S. Interface engineering of quantum Hall effects in digital transition metal oxide heterostructures. *Nature communications* **2**, 596 (2011).
7. Tsui, D. C., Stormer, H. L. & Gossard, A. C. Two-dimensional magnetotransport in the extreme quantum limit. *Physical Review Letters* **48**, 1559 (1982).
8. Laughlin, R. B. Anomalous quantum Hall effect: an incompressible quantum fluid with fractionally charged excitations. *Physical Review Letters* **50**, 1395 (1983).
9. Park, H. *et al.* Observation of fractionally quantized anomalous Hall effect. *Nature* **622**, 74-79 (2023).

10 Xu, F. *et al.* Observation of integer and fractional quantum anomalous Hall effects in twisted bilayer MoTe$_2$. *Physical Review X* **13**, 031037 (2023).
11 Cai, J. *et al.* Signatures of fractional quantum anomalous Hall states in twisted MoTe$_2$. *Nature* **622**, 63-68 (2023).
12 Zeng, Y. *et al.* Thermodynamic evidence of fractional Chern insulator in moiré MoTe$_2$. *Nature* **622**, 69-73 (2023).
13 Ji, Z. *et al.* Local probe of bulk and edge states in a fractional Chern insulator. *Nature* **635**, 578-583 (2024).
14 Redekop, E. *et al.* Direct magnetic imaging of fractional Chern insulators in twisted MoTe$_2$. *Nature* **635**, 584-589 (2024).
15 Park, H. *et al.* Ferromagnetism and Topology of the Higher Flat Band in a Fractional Chern Insulator. *arXiv preprint arXiv:2406.09591* (2024).
16 Xu, F. *et al.* Interplay between topology and correlations in the second moiré band of twisted bilayer MoTe$_2$. *arXiv preprint arXiv:2406.09687* (2024).
17 Lu, Z. *et al.* Fractional quantum anomalous Hall effect in multilayer graphene. *Nature* **626**, 759-764 (2024).
18 Lu, Z. *et al.* Extended quantum anomalous Hall states in graphene/hBN moiré superlattices. *Nature* **637**, 1090-1095 (2025).
19 Xie, J. *et al.* Tunable Fractional Chern Insulators in Rhombohedral Graphene Superlattices. *arXiv preprint arXiv:2405.16944* (2024).
20 Choi, Y. *et al.* Superconductivity and quantized anomalous Hall effect in rhombohedral graphene. *Nature*, 1-6 (2025).
21 Aronson, S. H. *et al.* Displacement field-controlled fractional Chern insulators and charge density waves in a graphene/hBN moiré superlattice. *arXiv preprint arXiv:2408.11220* (2024).
22 Jain, J. K. *Composite fermions*. (Cambridge University Press, 2007).
23 Halperin, B. I. & Jain, J. K. *Fractional quantum hall effects: new developments*. (World Scientific, 2020).
24 Chung, Y. J. *et al.* Ultra-high-quality two-dimensional electron systems. *Nature Materials* **20**, 632-637 (2021).
25 Anderson, E. *et al.* Programming correlated magnetic states with gate-controlled moiré geometry. *Science* **381**, 325-330 (2023).
26 Xia, Y. *et al.* Superconductivity in twisted bilayer WSe$_2$. *Nature* **637**, 833-838 (2024).
27 Knüppel, P. *et al.* Correlated states controlled by tunable van Hove singularity in moiré WSe$_2$. *arXiv preprint arXiv:2406.03315* (2024).
28 Dong, J., Wang, J., Ledwith, P. J., Vishwanath, A. & Parker, D. E. Composite Fermi liquid at zero magnetic field in twisted MoTe$_2$. *Physical Review Letters* **131**, 136502 (2023).
29 Goldman, H., Reddy, A. P., Paul, N. & Fu, L. Zero-field composite Fermi liquid in twisted semiconductor bilayers. *Physical Review Letters* **131**, 136501 (2023).
30 Anderson, E. *et al.* Magnetoelectric Control of Helical Light Emission in a Moiré Chern Magnet. *arXiv preprint arXiv:2503.02810* (2025).
31 Serlin, M. *et al.* Intrinsic quantized anomalous Hall effect in a moiré heterostructure. *Science* **367**, 900-903 (2020).
32 Li, T. *et al.* Quantum anomalous Hall effect from intertwined moiré bands. *Nature* **600**, 641-646 (2021).
33 Deng, Y. *et al.* Quantum anomalous Hall effect in intrinsic magnetic topological insulator MnBi$_2$Te$_4$. *Science* **367**, 895-900 (2020).

34 Schulze-Wischeler, F., Mariani, E., Hohls, F. & Haug, R. J. Direct measurement of the g factor of composite fermions. *Physical review letters* **92**, 156401 (2004).
35 Wang, M., Wang, X. & Vafek, O. Phase diagram of twisted bilayer $MoTe_2$ in a magnetic field with an account for the electron-electron interaction. *Physical Review B* **110**, L201107 (2024).
36 Zhang, S. *et al.* VASP2KP: k· p Models and Landé g-Factors from ab initio Calculations. *Chinese Physics Letters* **40**, 127101 (2023).
37 Halperin, B. I., Lee, P. A. & Read, N. Theory of the half-filled Landau level. *Physical Review B* **47**, 7312 (1993).
38 Stern, A. & Halperin, B. I. Singularities in the Fermi-liquid description of a partially filled Landau level and the energy gaps of fractional quantum Hall states. *Physical Review B* **52**, 5890 (1995).
39 Nayak, C. & Wilczek, F. Non-Fermi liquid fixed point in 2+ 1 dimensions. *Nuclear Physics B* **417**, 359-373 (1994).
40 Morf, R., d'Ambrumenil, N. & Sarma, S. D. Excitation gaps in fractional quantum Hall states: An exact diagonalization study. *Physical Review B* **66**, 075408 (2002).
41 Boebinger, G. *et al.* Activation energies and localization in the fractional quantum Hall effect. *Physical Review B* **36**, 7919 (1987).
42 Willett, R., Stormer, H., Tsui, D., Gossard, A. & English, J. Quantitative experimental test for the theoretical gap energies in the fractional quantum Hall effect. *Physical Review B* **37**, 8476 (1988).
43 Du, R., Stormer, H., Tsui, D., Pfeiffer, L. & West, K. Experimental evidence for new particles in the fractional quantum Hall effect. *Physical review letters* **70**, 2944 (1993).
44 Manoharan, H., Shayegan, M. & Klepper, S. Signatures of a novel fermi liquid in a two-dimensional composite particle metal. *Physical review letters* **73**, 3270 (1994).
45 Villegas Rosales, K. *et al.* Fractional quantum Hall effect energy gaps: Role of electron layer thickness. *Physical review letters* **127**, 056801 (2021).
46 Éfros, A. L. & Shklovskii, B. I. Coulomb gap and low temperature conductivity of disordered systems. *Journal of Physics C: Solid State Physics* **8**, L49 (1975).
47 Shklovskii, B. I., Efros, A. L., Shklovskii, B. I. & Efros, A. L. Variable-range hopping conduction. *Electronic properties of doped semiconductors*, 202-227 (1984).

**Acknowledgements:** The authors thank Jiaqi Cai, Shuai Yuan, and Yi-Fan Zhao for insightful discussion. The project was mainly supported by Programmable Quantum Materials, an Energy Frontier Research Center funded by DOE BES under award DE-SC0019443. K.W. and T.T. acknowledge support from the JSPS KAKENHI (Grant Numbers 20H00354, 21H05233 and 23H02052) and World Premier International Research Center Initiative (WPI), MEXT, Japan. B.A.B. was supported by the Gordon and Betty Moore Foundation (No.~GBMF8685 and No.~GBMF11070), ONR Grant No.~N00014-20-1-2303, the Simons Foundation Grants No.~404513 and SFI-MPS-NFS-00006741-01, the NSF-MERSEC (Grant No.~MERSEC DMR 2011750), the Schmidt Foundation at Princeton University, and European Research Council (ERC) No. 101020833. The Flatiron Institute is a division of the Simons Foundation.

**Author contributions:** XX conceived and supervised the project. HP fabricated the sample and performed transport measurements. WL performed the RMCD measurements with support from CB. HP, DHC, LF, BAB, NR, DX, and XX analyzed and interpreted the results. MG, JMV, JHA,

DX, BAB, NR performed the magnetic field dependent gap analysis. TT and KW synthesized the hBN crystals. CH and JHC grew and characterized the bulk $MoTe_2$ crystals. HP, DX, and XX wrote the paper with input from all authors. All authors discussed the results.

**Competing Interests:** The authors declare no competing interests.

**Data Availability:** The datasets generated during and/or analyzed during this study are available from the corresponding author upon reasonable request.

# Figures:

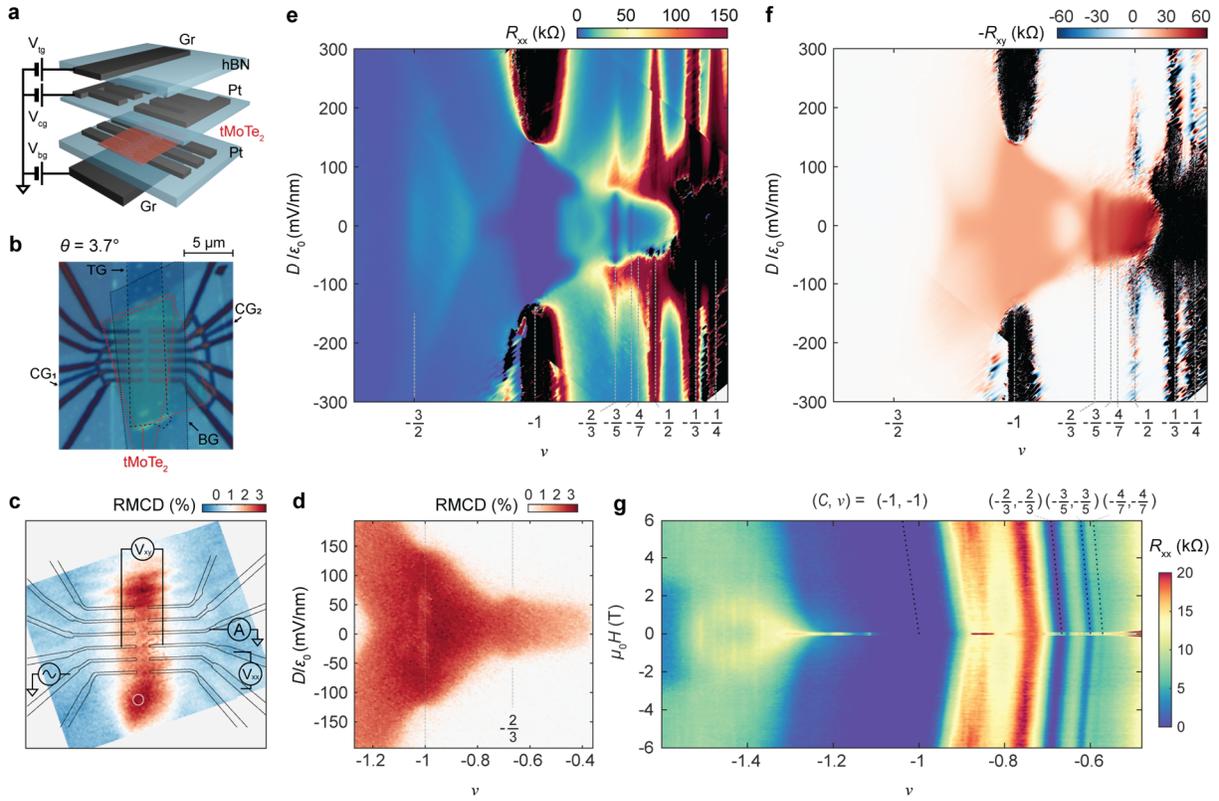

**Figure 1 | Fractional Chern insulator (FCI) phase diagram of the 3.7° twisted MoTe₂ (tMoTe₂) device. a,** A schematic of the device illustrating the contact gate design. A dual-gate configuration is formed by graphite (Gr) top ($V_{tg}$) and bottom ($V_{bg}$) gates, while patterned platinum (Pt) local gates ($V_{cg}$) are incorporated to improve contact resistance. **b,** Optical image of the device. Boundaries of each MoTe₂ flake is denoted in red dotted lines and the top gate (TG) defines the edges of the device. **c,** Spatial map of reflective magnetic circular dichroism (RMCD) at filling factor $v = -1$ and zero electric field, showing a clear enhancement of magnetization at the region defined by the top gate. The transport measurement configuration is illustrated on top. **d,** Magnetic circular dichroism signal versus $v$ and $D/\varepsilon_0$. The data is taken at 1.6 K and a magnetic field of 5 mT at the position denoted by the white circle in **c**. **e,** Longitudinal resistance $R_{xx}$, and **f,** transverse resistance $-R_{xy}$ as a function of filling factor ($v$) and applied electric field ($D/\varepsilon_0$). The data is taken at base temperature of 10 mK. $R_{xx}$ and $-R_{xy}$ are symmetrized and antisymmetrized at an external magnetic field of $\mu_0 H = \pm 0.2$ T. **g,** Fan diagram of $R_{xx}$ as a function of magnetic field and doping (or filling factor $v$). Both integer ($C = -1$) and fractional ($C = -2/3, -3/5,$ and $-4/7$) Chern insulator states are identified. Dashed lines are overlaid carrier density – magnetic field dependence for the Chern insulator states from the Streda formula.

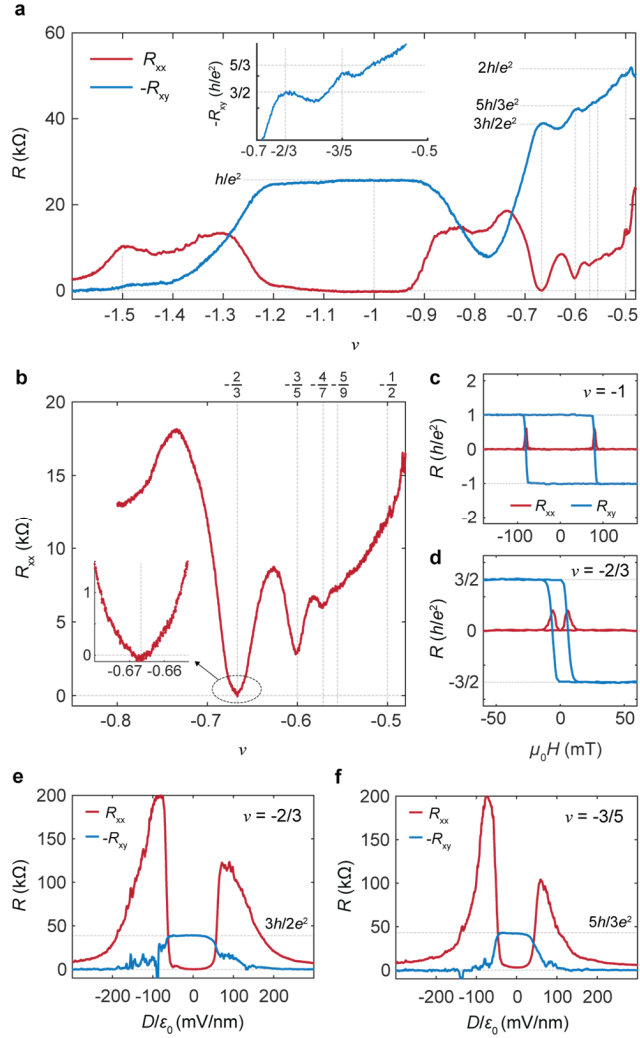

**Figure 2 | Observation of dissipationless FCI. a,** Symmetrized $R_{xx}$ (red) and antisymmetrized $-R_{xy}$ (blue) at $\mu_0 H = \pm 100$ mT as a function of filling factor $v$. The data is taken at base temperature of 10 mK. The quantum anomalous Hall state at $v = -1$, and fractional Chern insulator states along the Jain sequence $-p/(2p-1)$ ($p = 2, 3, 4, 5$) are observed, evident by the quantized $R_{xy}$ (inset) and $R_{xx}$ dips. **b,** Fine doping dependent $R_{xx}$ across the fractional fillings, emphasizing the pronounced $R_{xx}$ dips at the FCI states. (Inset) Zoom in plot near $v = -2/3$ which shows that $R_{xx}$ approaches zero for the FCI state. **c, d,** $R_{xx}$ (red), $R_{xy}$ (blue) versus cycling magnetic field at the quantum anomalous Hall ($v = -1$, **c**) and fractional Chern insulator ($v = -2/3$, **d**) states, respectively. **e, f,** Electrical field ($D/\varepsilon_0$) dependent $R_{xx}$ and $-R_{xy}$ at $v = -2/3$ and $-3/5$, respectively. An extended range of quantization can be seen near zero electric field (see Extended Data Fig. 5 for details).

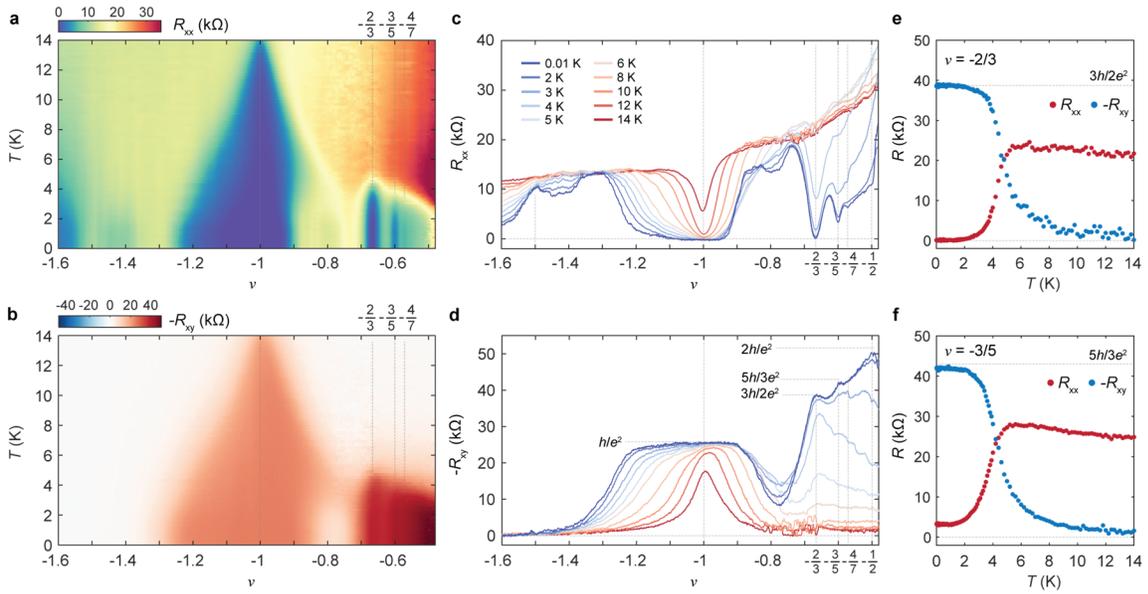

**Figure 3 | Temperature dependent measurements. a,** $R_{xx}$ and **b,** $-R_{xy}$ as a function of filling factor ($v$) and temperature ($T$) from 10 mK to 14 K. The data are symmetrized and antisymmetrized at $\mu_0 H = \pm 150$ mT. **c,** Linecuts of $R_{xx}$ at selected temperatures from panel (**a**). **d,** Linecuts of $R_{xy}$ at selected temperature from panel (**b**). **e,** Temperature dependence of $R_{xx}$ (red) and $-R_{xy}$ (blue) at the fixed filling of $v = -2/3$. **f,** Temperature dependence of $R_{xx}$ and $-R_{xy}$ at the fixed filling of $v = -3/5$.

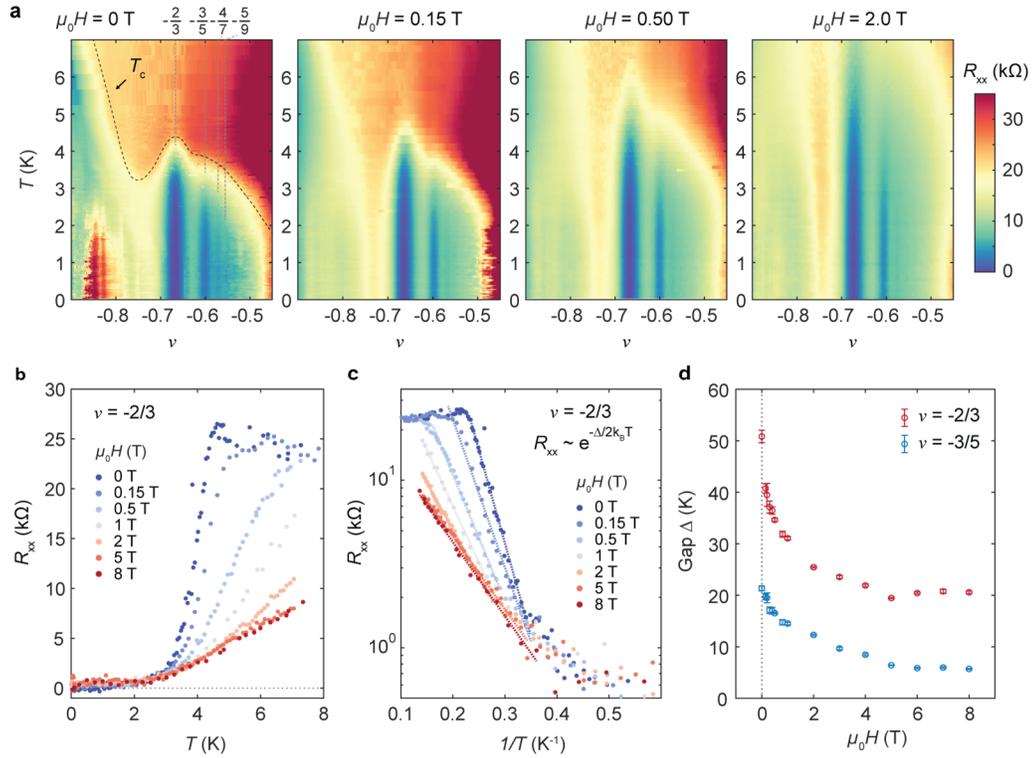

**Figure 4 | Magnetic field dependent FCI gap measurements. a**, $R_{xx}$ versus temperature ($T$) and filling factor ($v$) at selected magnetic fields ($\mu_0 H$). From left to right, $\mu_0 H$ = 0, 0.15 T, 0.5 T, 2 T. Dashed lines in panel (a) indicates Curie temperature $T_C$, determined from magnetic circular dichroism measurements (Extended Data Fig. 8, 9). **b**, Extracted $R_{xx}$ versus temperature of the -2/3 state at selected magnetic fields. **c**, Arrhenius plots of the -2/3 state at selected magnetic fields. Dashed line corresponds to the thermal activation fit to the equation $R_{xx} \sim \exp(-\Delta/2k_B T)$. The reduced slope as magnetic field increases are clearly observed. **c**, Extracted thermal activation gap $\Delta$ as a function of magnetic field for the -2/3 (red) and -3/5 (blue) state. Error bars correspond to the fitting uncertainty.

# Extended Data Figures for

## Observation of High-Temperature Dissipationless Fractional Chern Insulator

**Authors:** Heonjoon Park[1], Weijie Li[1], Chaowei Hu[1], Christiano Beach[1], Miguel Gonçalves[2], Juan Felipe Mendez-Valderrama[2], Jonah Herzog-Arbeitman[2], Takashi Taniguchi[3], Kenji Watanabe[4], David Cobden[1], Liang Fu[5], B. Andrei Bernevig[2,6,7], Nicolas Regnault[8,2,9], Jiun-Haw Chu[1], Di Xiao[10,1], Xiaodong Xu[1,10*]

[1]Department of Physics, University of Washington, Seattle, Washington 98195, USA
[2]Department of Physics, Princeton University, Princeton, New Jersey 08544, USA
[3]Research Center for Materials Nanoarchitectonics, National Institute for Materials Science, 1-1 Namiki, Tsukuba 305-0044, Japan
[4]Research Center for Electronic and Optical Materials, National Institute for Materials Science, 1-1 Namiki, Tsukuba 305-0044, Japan
[5]Department of Physics, Massachusetts Institute of Technology, Cambridge, Massachusetts 02139, USA
[6]Donostia International Physics Center, P. Manuel de Lardizabal 4, 20018 Donostia-San Sebastian, Spain
[7]IKERBASQUE, Basque Foundation for Science, Bilbao, Spain
[8]Center for Computational Quantum Physics, Flatiron Institute, 162 5th Avenue, New York, NY 10010, USA
[9]Laboratoire de Physique de l'Ecole Normale Supérieure, ENS, Université PSL, CNRS, Sorbonne Université, Université Paris-Diderot, Sorbonne Paris Cité, 75005 Paris, France
[10]Department of Materials Science and Engineering, University of Washington, Seattle, Washington 98195, USA
*Corresponding author's email: xuxd@uw.edu

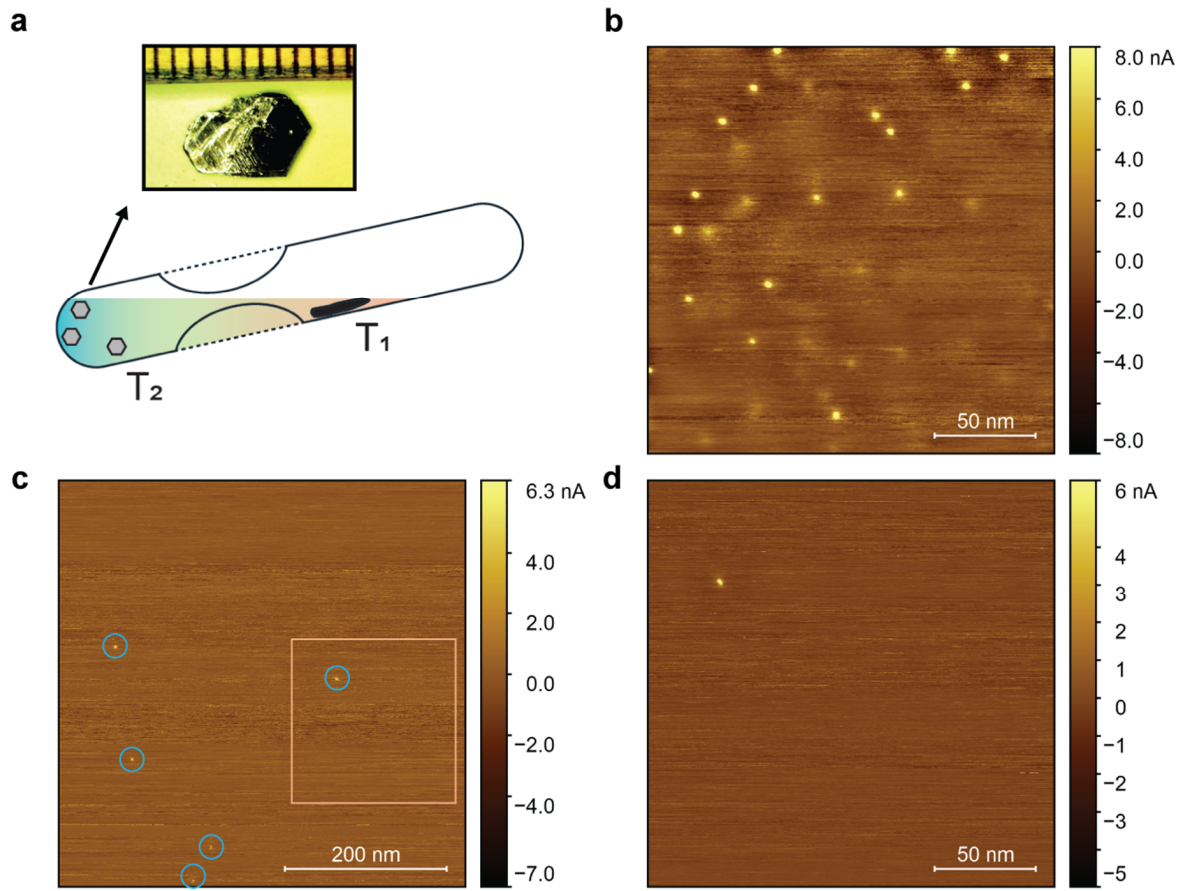

**Extended Data Figure 1 | Horizontal flux growth and characterizations of bulk MoTe2 crystals. a**, Schematic of horizontal flux transport growth of MoTe$_2$ crystal in Te flux under temperature gradient from T$_1$=600 °C to T$_2$=550 °C. **b**, 200×200 nm$^2$ conductive atomic force microscopy (c-AFM) image taken on exfoliated bulk commercial crystals under V$_{bias}$= 1.3 V showing several types of defects with the total concentration on the order of 10$^{11}$ cm$^{-2}$. The exact defect types are not known. **c,** 500×500 nm$^2$ c-AFM image taken on exfoliated bulk crystals grown by horizontal flux transport (see Methods). The applied V$_{bias}$= 0.9 V. A single type of defects is highlighted in blue circles with density of 2× 10$^9$ cm$^{-2}$. **d,** 200×200 nm$^2$ c-AFM image taken within the orange box in c.

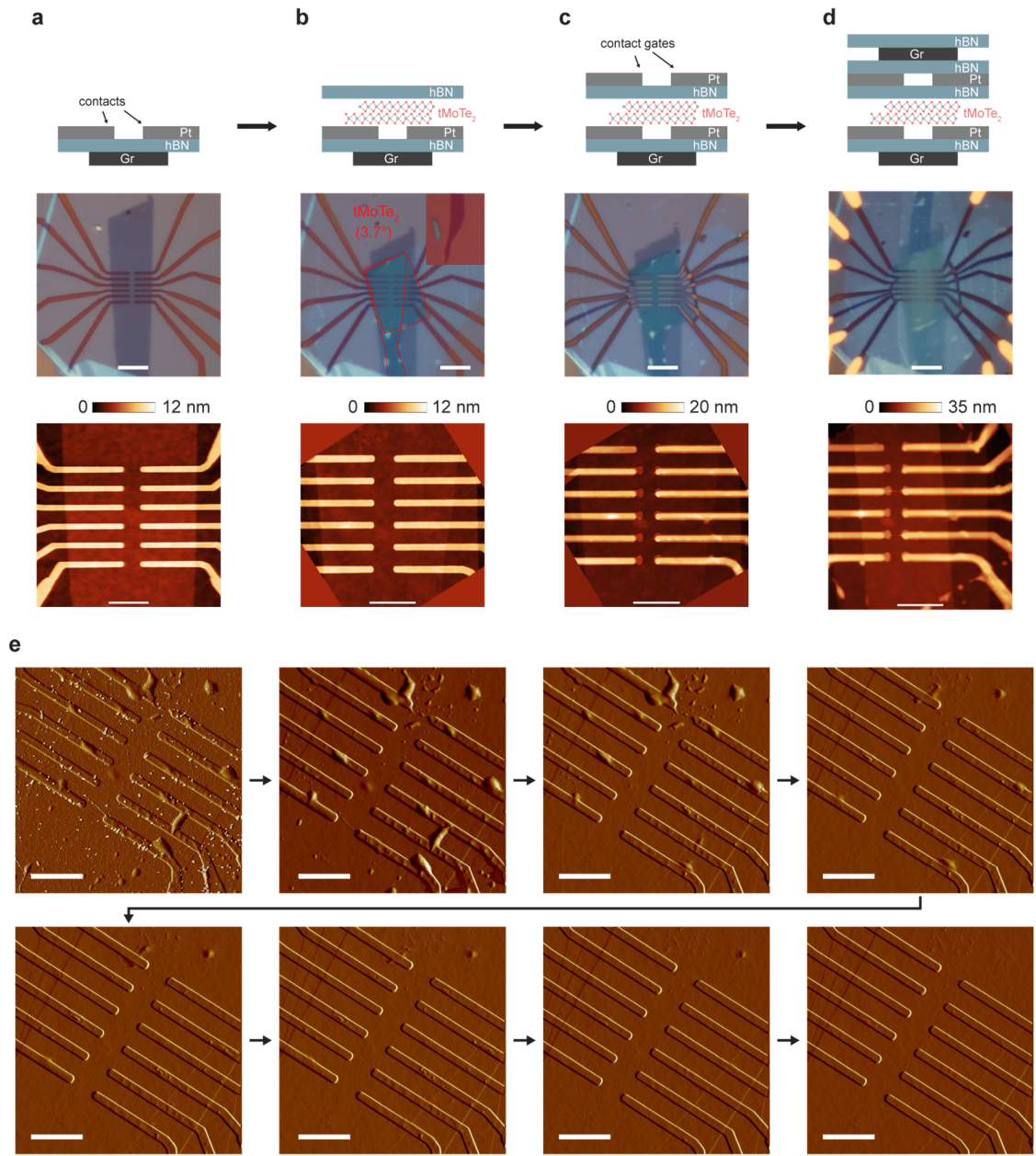

**Extended Data Figure 2 | Device fabrication. a-d,** Each row displays, from top to bottom, the device schematic, optical image, and atomic force microscopy (AFM) image. **a,** Bottom gate structure featuring pre-patterned platinum (Pt) electrodes on an hBN/graphite stack. **b,** Placement of the hBN/tMoTe$_2$ stack onto the bottom gate. Inset: Optical image of a monolayer MoTe$_2$ flake bisected using an AFM tip. **c,** Patterning of Pt contact gates on top of each Pt electrode. **d,** Transfer of the top gate structure (hBN/graphite/hBN) onto the device. **e,** Contact-mode AFM images taken after hBN/tMoTe$_2$ transfer in **b**, illustrating the progressive removal of bubbles outside the Hall bar region. Scale bars: 5 μm (optical images), 2 μm (AFM images).

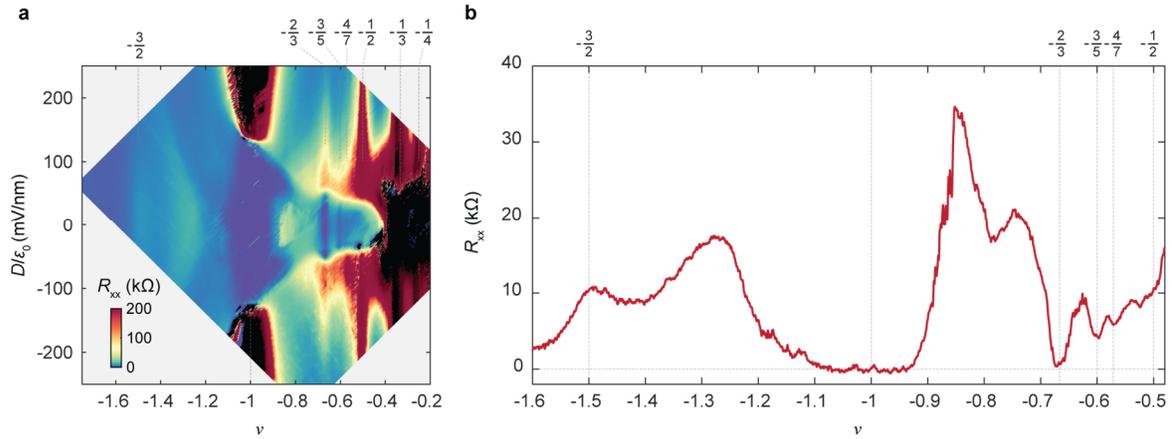

**Extended Data Figure 3 | Zero magnetic field phase diagram. a,** Longitudinal ($R_{xx}$) resistance as a function of filling factor ($v$) and electric field ($D/\varepsilon_0$) at 10 mK. **b,** Linecut of $R_{xx}$ as a function of filling factor at $D/\varepsilon_0 = 25$ mV/nm. The $R_{xx}$ is symmetrized at $|\mu_0 H| = \pm 0$ mT by initializing the device at $\pm 0.2$ T and sweeping down the magnetic field to zero. The observed phase diagram is the same as the one shown in Fig. 1c, except small fluctuations in $R_{xx}$ due to spin fluctuations during carrier density sweeps.

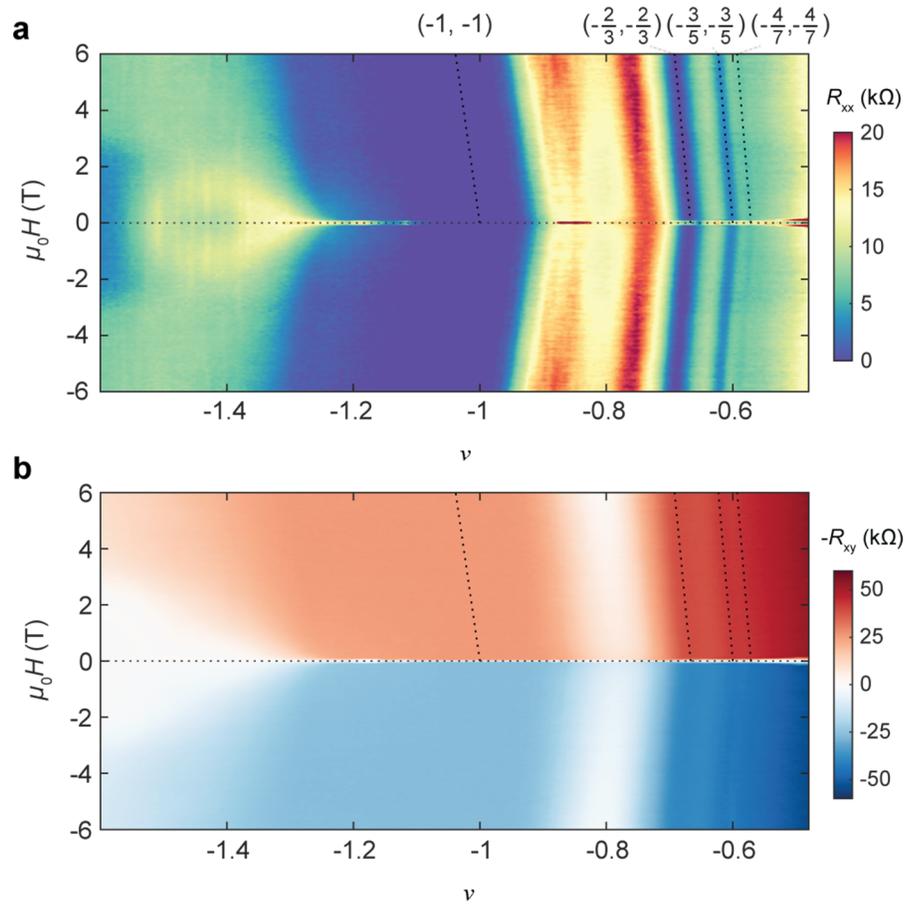

**Extended Data Figure 4 | Fan diagram and Streda Formula analysis. a,** $R_{xx}$ and **b,** $-R_{xy}$ as a function of magnetic field ($\mu_0H$) and filling factor ($\nu$). Data is taken at base temperature of 10 mK and symmetrized and antisymmetrized, respectively. The overlaid dashed lines are expected carrier density and magnetic field dependence for corresponding Chern insulator states ($C, \nu$), where $C$ is the Chern number and $\nu$ is the filling factor of the first moiré Chern band.

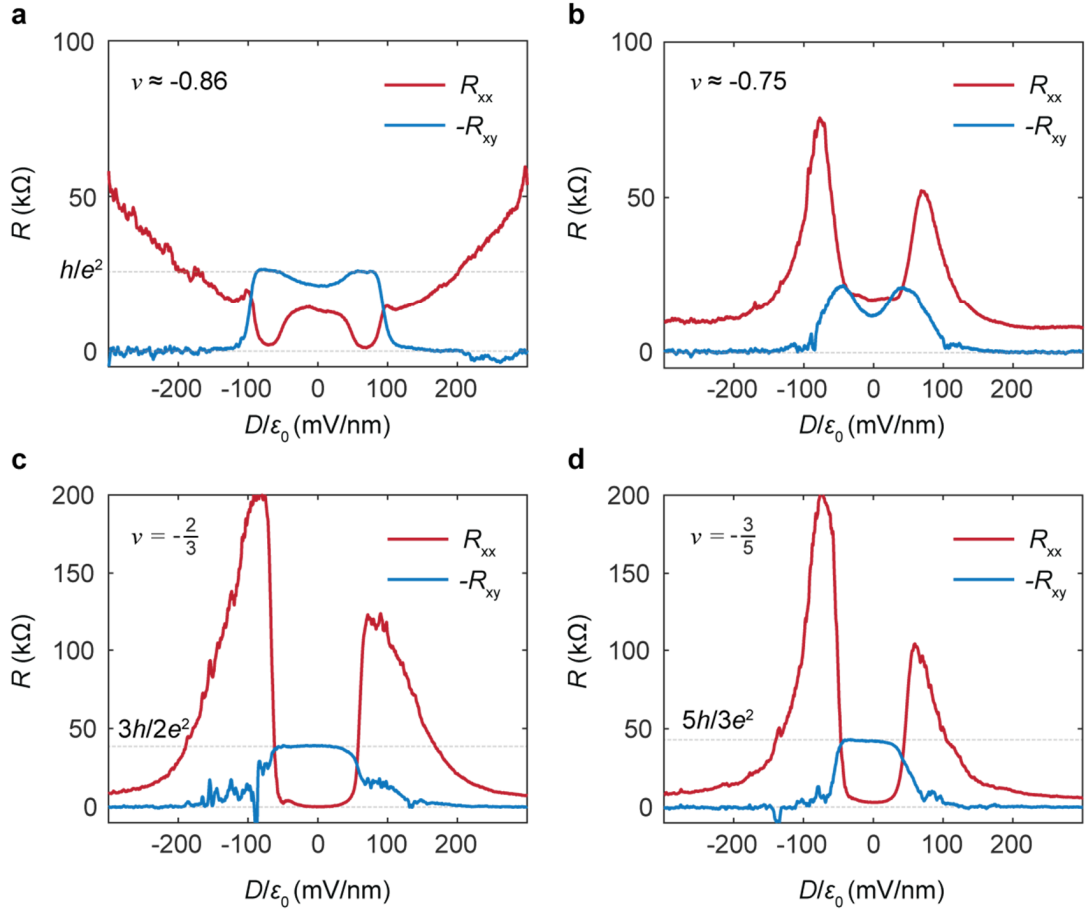

**Extended Data Figure 5 | Electrical field-dependent phase transitions at selected filling factors. a-d,** Plots of $R_{xx}$ (red) and $-R_{xy}$ (blue) versus electric field $D/\varepsilon_0$ at various filling factors. The data are symmetrized and antisymmetrized at $\mu_0 H = \pm 100$mT. **a**, At $\nu \approx -0.86$, the system first shows a weakly resistive state with large $-R_{xy}$ at small $D/\varepsilon_0$. As $D/\varepsilon_0$ increases to about 75 mV/nm, a transition to a quantum anomalous Hall state with vanishing $R_{xx}$ and quantized $-R_{xy} \approx h/e^2$ occurs, followed by a topologically trivial state with full layer polarization at larger $D/\varepsilon_0$. **b**, At $\nu \approx -0.75$, $R_{xx}$ slightly exceeds $-R_{xy}$ for $|D/\varepsilon_0| < 50$ mV/nm. Increasing $D/\varepsilon_0$ toward full layer polarization causes $R_{xx}$ to rise sharply while $-R_{xy}$ decreases. Beyond this point, $R_{xx}$ drops with vanishing $-R_{xy}$, indicating a topologically trivial metallic phase. **c**, For the $\nu = -2/3$ FCI state, $-R_{xy}$ is quantized and $R_{xx}$ is zero at low $D/\varepsilon_0$. Once $D/\varepsilon_0$ passes approximately 50 mV/nm, $R_{xx}$ quickly increases while $-R_{xy}$ decreases, signifying a transition from the FCI to a charge-ordered state. As $D/\varepsilon_0$ continues to increase, $R_{xx}$ drops again, pointing to a metallic state. **d**, the $\nu = -3/5$ FCI state shows a similar progression with $D/\varepsilon_0$. The main difference is that after $R_{xx}$ reaches its maximum, it decreases more sharply than in the $\nu = -2/3$ state as $D/\varepsilon_0$ increases. Panels (c) and (d) are reproduced from Figs. 2e and f in the main text.

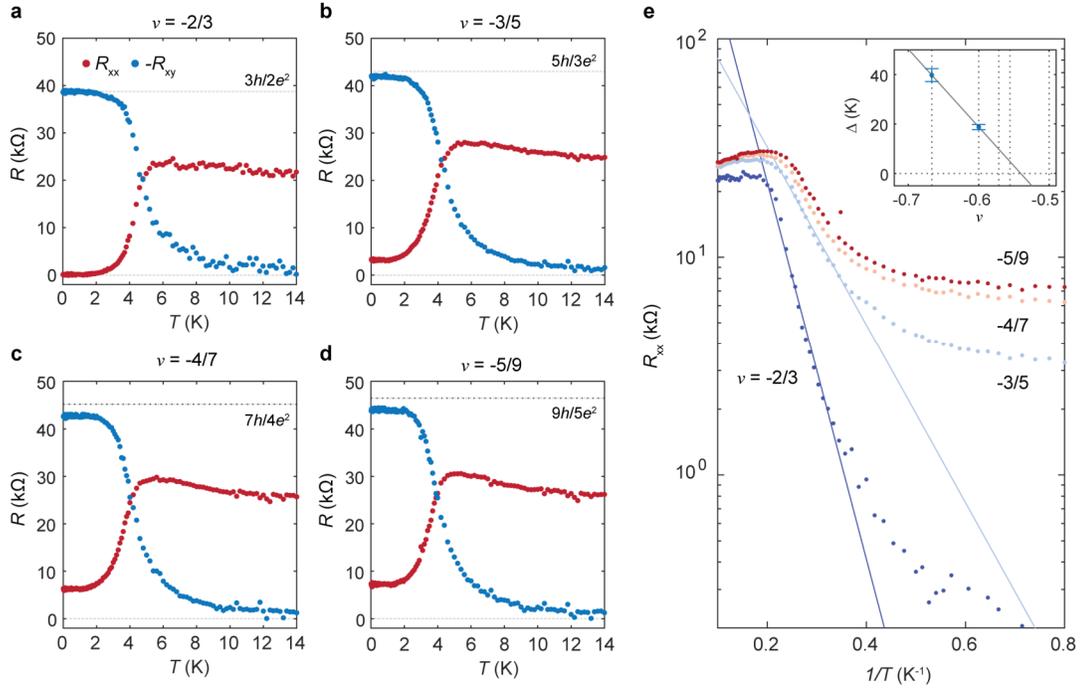

**Extended Data Figure 6 | Thermal activation gap fitting. a,** Symmetrized $R_{xx}$ and antisymmetrized $-R_{xy}$ at $\mu_0 H = \pm 150$ mT for the -2/3 FCI state. $R_{xx} \approx -R_{xy}$ around a temperature of 5 K. **b**, **c**, and **d**, similar measurements as (**a**) for the -3/5, -4/7, and -5/9 FCI state. **e,** Arrhenius plot of $R_{xx}$ for the -2/3, -3/5, -4/7, and -5/9. The extracted thermal activation gap is shown in the inset for the -2/3 and -3/5 state, where the line is a guide to the eye for the expected scaling behavior for the fractional quantum Hall states as a reference.

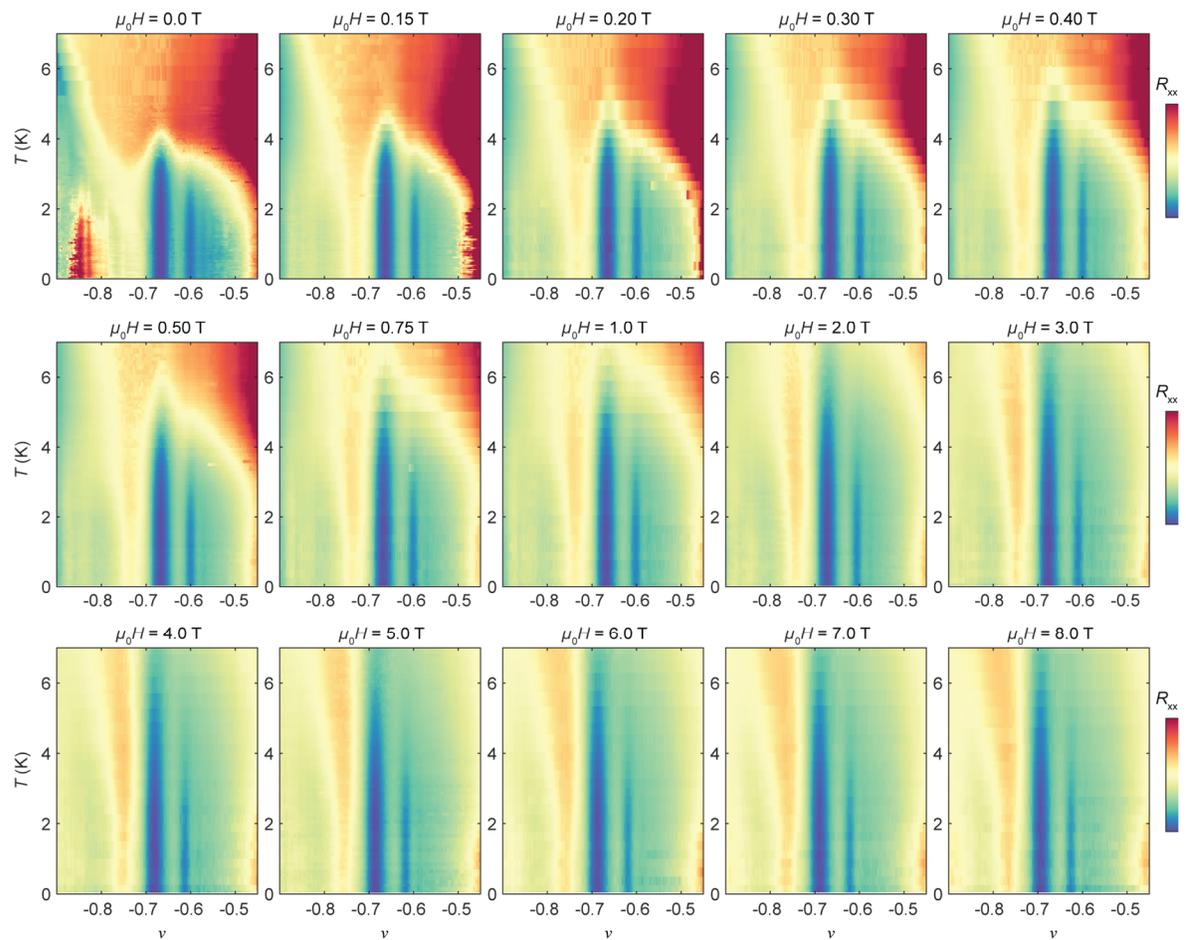

**Extended Data Figure 7 | Temperature dependent $R_{xx}$ versus filling factors at selected magnetic fields.** For each panel, the corresponding magnetic field is marked on the top.

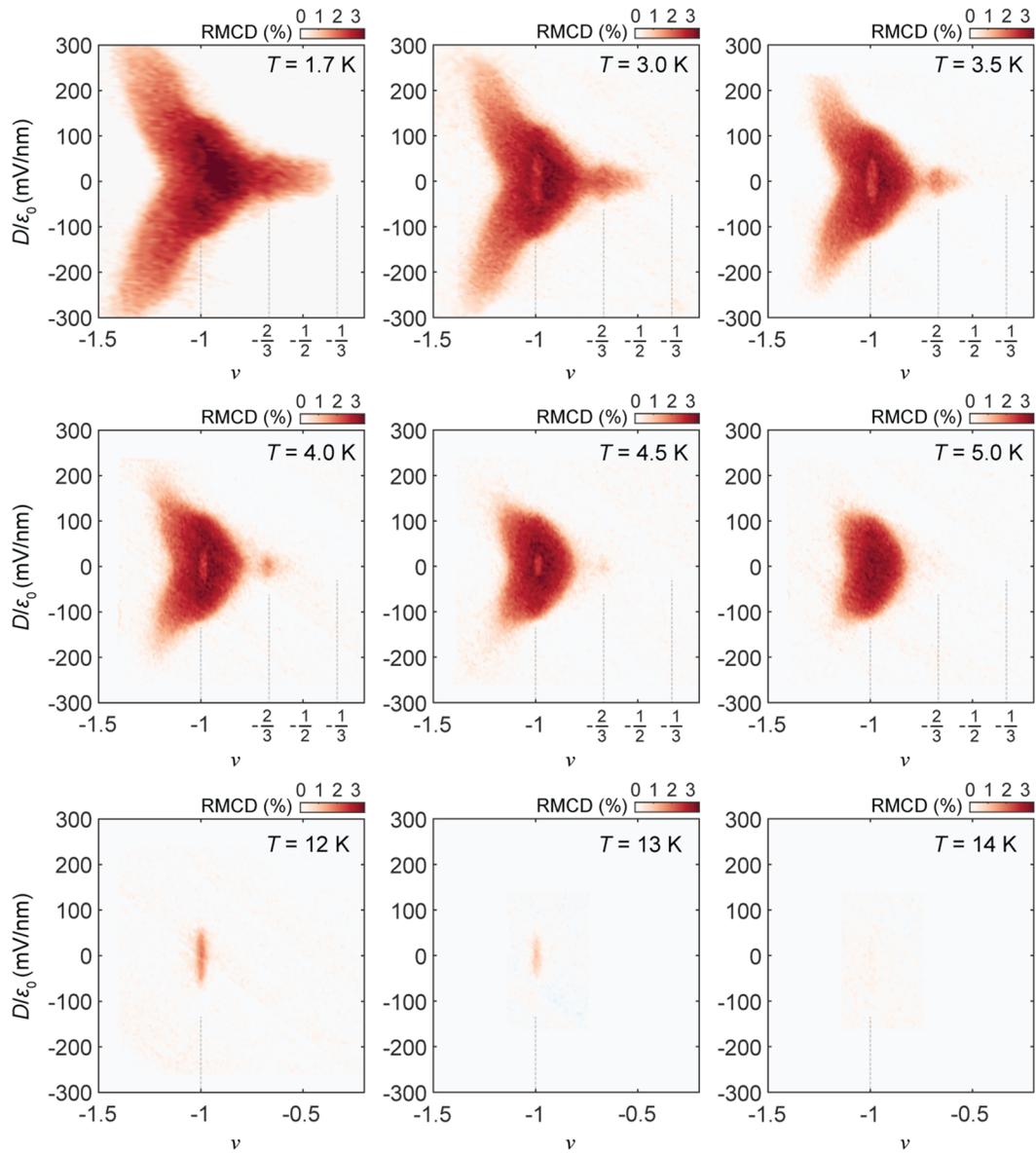

**Extended Data Figure 8 | Temperature dependent RMCD versus filling factors.** RMCD as a function of $v$ and $D/\varepsilon_0$ at different temperatures ($T$), measured at a small field of 5 mT. The ferromagnetic signal at $v = -2/3$ disappears around 4-5 K while the $v = -1$ signal persists up to approximately 13-14 K, consistent with the temperature dependence of $R_{xy}$.

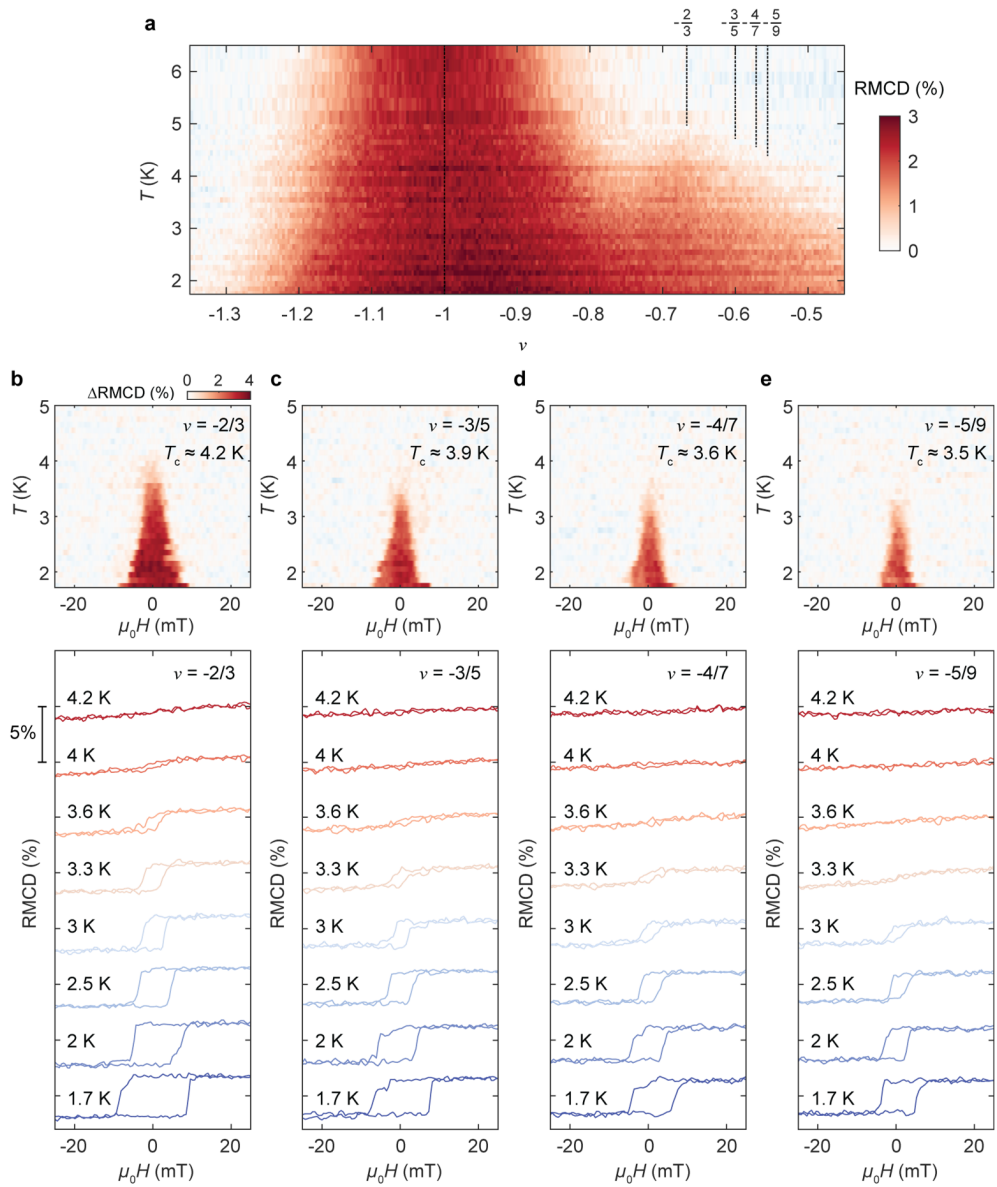

**Extended Data Figure 9 | Determination of Curie temperature**. **a**, RMCD as a function of $v$ at different temperatures ($T$), measured at a small field of 30 mT. The features are qualitatively similar to the $R_{xy}$ seen in Fig. 3b. **b-e**, (top) The hysteretic component of RMCD (ΔRMCD) as a function of magnetic field and temperature for the Jain sequence fractional fillings $v = -2/3, -3/5, -4/7, -5/9$, respectively. (bottom) Linecuts at different temperatures extracted from the colorplot. The Curie temperature decreases at higher order filling factors.

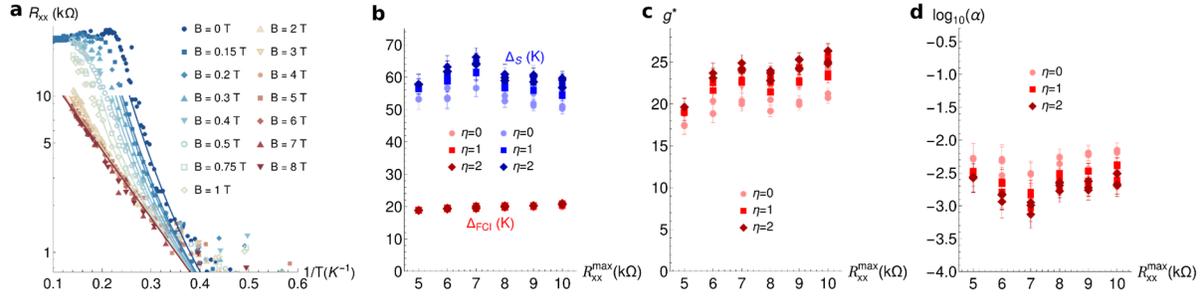

**Extended Data Figure 10 | Fits of the experimental data to Eqn.(3).** a, Raw data for the longitudinal resistance $R_{xx}$ at different temperatures and magnetic fields, together with an example fit (solid line) to Eq.3 by selecting the data for $0.15 K^{-1} < T^{-1} < 0.35 K^{-1}$ and $R_{xx} < 7.5 k\Omega$. The different magnetic fields are color-coded according to the legend. The extracted fitted parameters are $\Delta_{FCI} = 19.9 \pm 0.3 K$, $\Delta_S^0 = 55 \pm 2 K$, $g^* = 19.3 \pm 0.8$ and $\alpha = (4 \pm 1) \times 10^{-3}$. b-d, Best-fit parameters for different choices of data selection. These include different choices of truncation ranging from $0.1 K^{-1} < T_{max}^{-1} < 0.2 K^{-1}$ (always considering $T^{-1} < 0.35 K^{-1}$) and $5 k\Omega < R_{xx}^{max} < 10 k\Omega$, where $T_{max}$ and $R_{xx}^{max}$ are respectively the maximum considered temperatures and resistances. We also tested the impact of fitting with larger weight in the small $B$ data, where the most significant variations occur. We used different weighting functions $f(B) = ((B_{max} - B)/B_{max})^\eta$, with $\eta = 0,1,2$ indicated by the different colors of the data points in the figure, and $B_{max} = 8T$.